\documentclass[showpacs,amsmath,amssymb,aps,showkeys,floatfix,prd,a4paper]{revtex4}

\usepackage[dvips]{graphicx}
\usepackage{dcolumn}
\usepackage{bm}
\usepackage{epsfig}
\usepackage{amsfonts}
\usepackage{amssymb,amscd}

\def\lsim{\raise0.3ex\hbox{$<$\kern-0.75em\raise-1.1ex\hbox{$\sim$}}}
\def\gsim{\raise0.3ex\hbox{$>$\kern-0.75em\raise-1.1ex\hbox{$\sim$}}}

\def\beq{\begin{equation}}
\def\eeq{\end{equation}}
\def\bea{\begin{eqnarray}}
\def\eea{\end{eqnarray}}
\def\bq{\begin{quote}}
\def\eq{\end{quote}}

\newcommand{\rr}{\mbox{\boldmath $r$}}

\newcommand{\rb}{\mbox{\boldmath $b$}}

\def\gappeq{\mathrel{\rlap {\raise.5ex\hbox{$>$}}
{\lower.5ex\hbox{$\sim$}}}}

\def\lappeq{\mathrel{\rlap{\raise.5ex\hbox{$<$}}
{\lower.5ex\hbox{$\sim$}}}}

\def\Toprel#1\over#2{\mathrel{\mathop{#2}\limits^{#1}}}

\begin{document}

%\preprint{Version 1.2}

\title{Exclusive heavy quark photoproduction in $pp$, $pPb$ and $PbPb$ collisions at the LHC and FCC energies}

\author{V.~P. Gon\c{c}alves, G. Sampaio dos Santos, C. R. Sena }
%\email{barros@ufpel.edu.br}
\affiliation{High and Medium Energy Group, \\
Instituto de F\'{\i}sica e Matem\'atica, Universidade Federal de Pelotas\\
Caixa Postal 354, CEP 96010-900, Pelotas, RS, Brazil }

%\\
%$^4$ Institut de Physique Th\'eorique, Universit\'e Paris Saclay,\\ 
%CEA, CNRS, F-91191, Gif-sur-Ivette, France\\

\date{\today}

\begin{abstract}
In this paper we present a comprehensive analysis of  the exclusive heavy quark photoproduction in $pp$, $pPb$ and $PbPb$ collisions at LHC and FCC energies using the color dipole formalism and taking into account of nonlinear corrections to the QCD dynamics. We estimate the rapidity distributions and  cross sections for the charm and bottom production   considering  three phenomenological models for the dipole-proton scattering amplitude that are able to describe the  $ep$ HERA data. 
Our results indicate that a future experimental analysis of this process is feasible, which will allow us to improve our understanding of the QCD dynamics.

\end{abstract}

\maketitle

%%%%%%%%%%%%%%%%%%%%%%%%%%%%%%%%%%%%%%%%%%%%%%%%%%%%%%%%%%%%%%%%%%%%%%%%%%%%%%%%

\section{Introduction}

One of the main goals of Particle Physics is to achieve a deeper knowledge of the hadronic structure.
A multidimensional partonic imaging of the hadron is provided by the 5-dimensional QCD Wigner distributions, which encode all quantum information about partons, including information on both generalized parton distributions (GPD) and transverse momentum dependent parton distributions (TMD) (See, e.g Refs. \cite{Ji:2003ak,Diehl:2003ny,Boer:2011fh,Hagiwara:2016kam}). 
In the last years several authors proposed  to constrain  the gluon Wigner distribution in the nucleon by studying different final states that can be generated in photon-induced interactions present in electron-hadron and hadron-hadron collisions \cite{Hatta:2016dxp,Altinoluk:2015dpi,Hagiwara:2017fye,Boussarie:2018zwg,Mantysaari:2019csc,Salazar:2019ncp,Boussarie:2019ero,roman_emmanuel,Hatta:2019ixj}. One of the more promising processes is the exclusive dijet photoproduction in ultraperipheral hadronic collisions \cite{Hagiwara:2017fye}, which are characterized by an impact parameter that is larger than the sum of the radius of the incident hadrons \cite{upc}. In such collisions the final state is very clean, being  characterized by the dijet, two intact hadrons and two rapidity gaps associated to the  photon and Pomeron exchanges. However, the measurement of the angular distribution of the Wigner distribution in this final state is challenging, since it requires reconstruction of full dijet kinematics. An alternative is to  consider the exclusive heavy quark photoproduction in hadronic collisions \cite{roman_emmanuel}. As demonstrated for the first time in Ref. \cite{vicmag_hqdif}, such process probes the nonlinear corrections to the QCD dynamics at high energies \cite{hdqcd}. In order to reconstruct the isotropic and elliptic components of the  gluon Wigner distribution, it is fundamental to access the dependence of the differential distribution in the relative quark-antiquark momentum for distinct values of the momentum transfer. Such analysis will only be feasible if the corresponding number of events generated in the current and/or future colliders is large.
The main goal of this paper is to estimate the cross sections for the exclusive charm and bottom photoproduction in $pp$, $pPb$ and $PbPb$ collisions using the Color Glass Condensate formalism \cite{CGC}. In our study we will consider two very successfull implementations of this formalism, the bCGC and IP-SAT models, which are able to describe the inclusive and exclusive $ep$ HERA data.  For comparison,  predictions derived using the IPnonSAT model, which disregards the nonlinear effects, will also presented.  We will derive  predictions for the rapidity distributions and cross sections 
for the LHC energies, which update the results presented in Ref. \cite{vicmag_hqdif}. Moreover, predictions for  the center-of-mass energies of the Future Circular Collider (FCC) \cite{fcc} will be presented for the first time. Finally, a comparison between the predictions for the exclusive and inclusive heavy quark photoproduction  will also be presented. As we will demonstrate below, our results indicate that a future experimental analysis of the exclusive heavy quark photoproduction is feasible and that this process can be used to improve our understanding of the QCD dynamics. In addition, the large number of events expected for the LHC and FCC, in particular for charm production, will allow to study more differential distributions, as those necessary to constrain the elliptic component of the gluon Wigner function.

 The paper is organized as follows. In  Sec. \ref{form} we present a brief review of the color dipole formalism and the main expressions 
used to estimate the   exclusive heavy quark photoproduction. Moreover, the distinct models for the dipole -- hadron scattering amplitude are discussed.  In Sec. \ref{res}, we present our predictions for the cross 
sections and 
rapidity distributions to be measured in $pp/pPb/PbPb$ collisions at the  LHC and FCC energies. 
Finally, in Sec. \ref{conc}, we summarize our main conclusions.

\begin{figure}[t]
\begin{tabular}{cc}
\centerline{{\includegraphics[height=6cm]{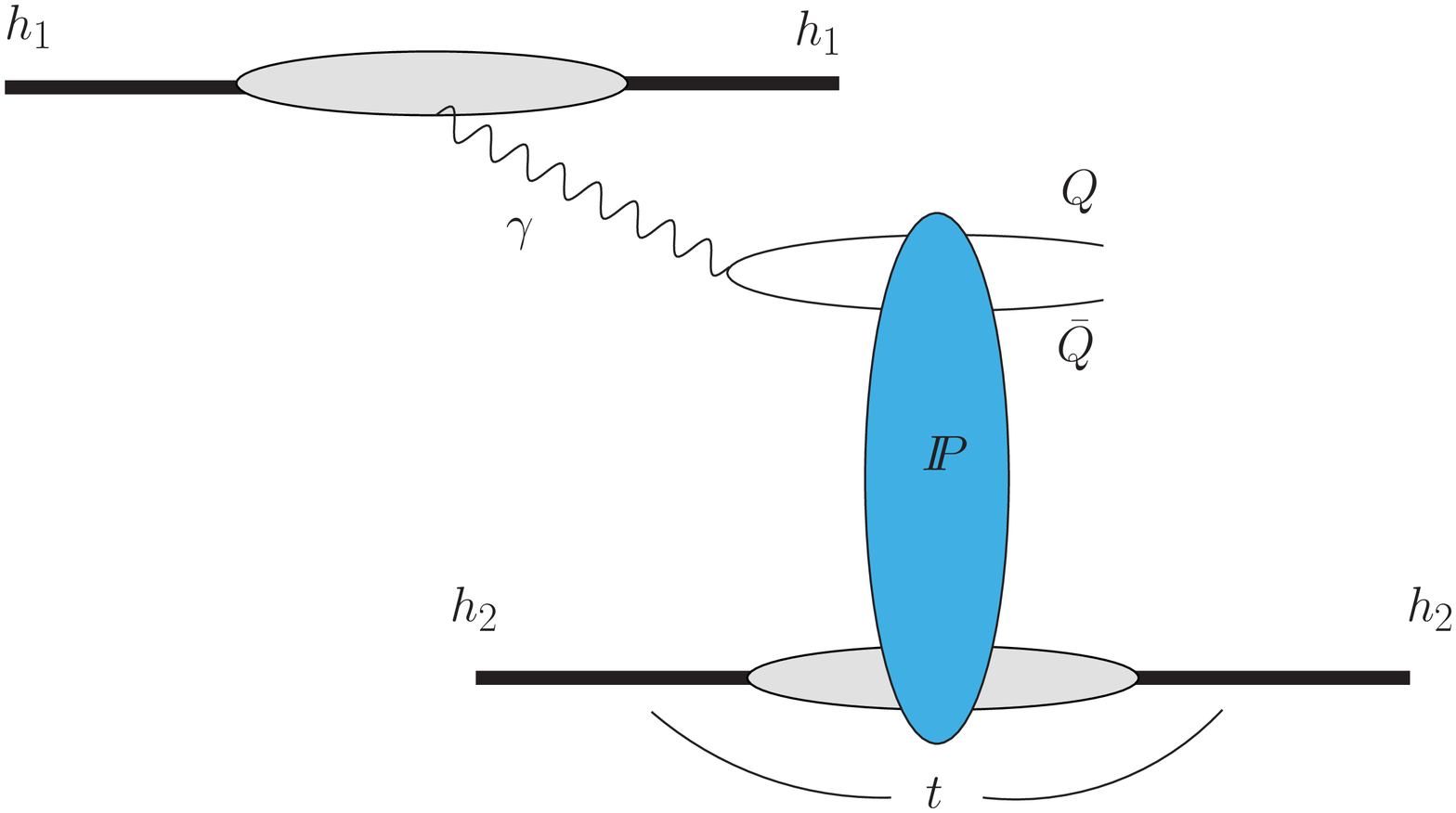}}
{\includegraphics[height=6cm]{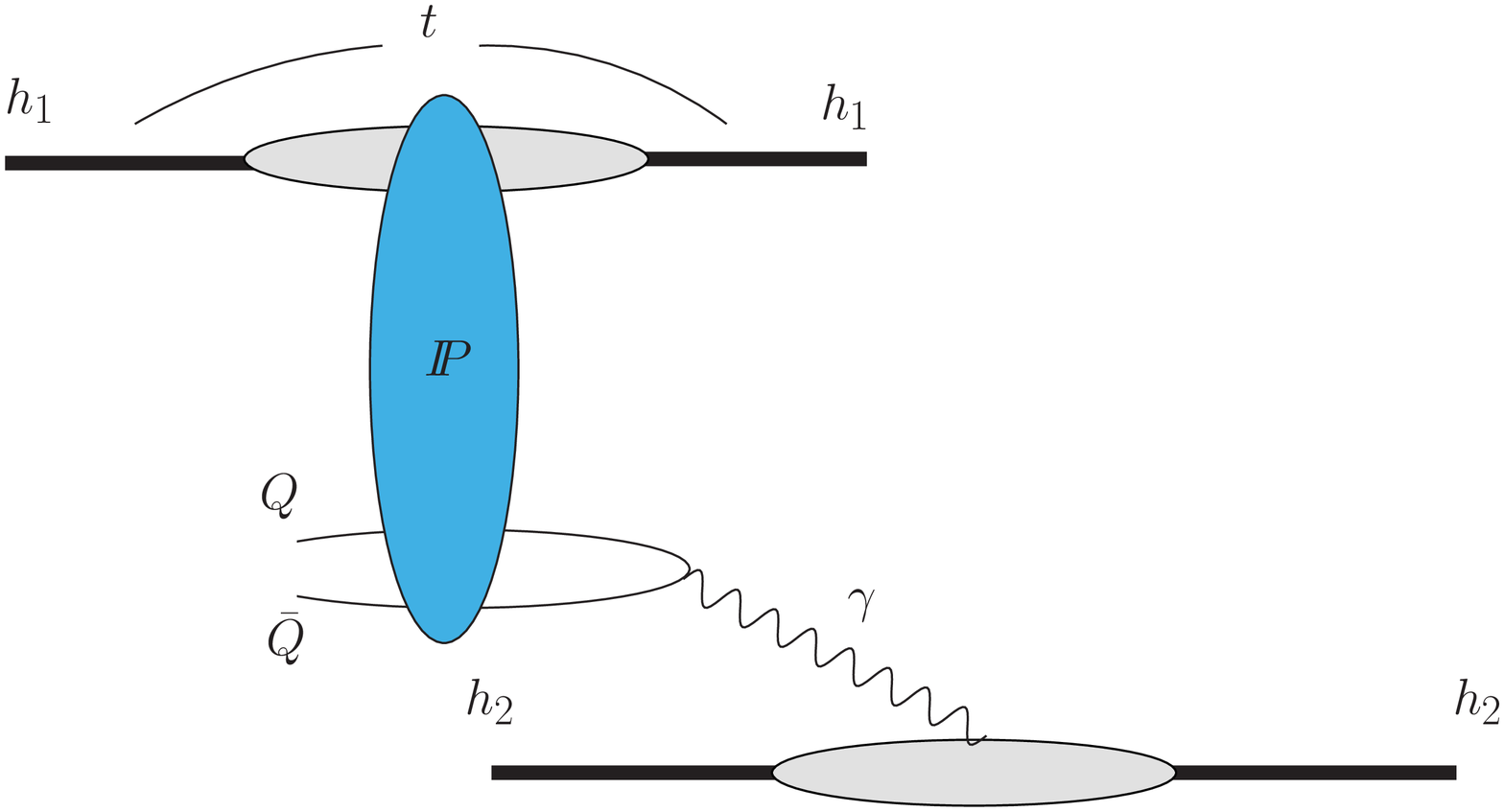}}}
\end{tabular}
\caption{Typical diagrams for the exclusive heavy quark photoproduction in a hadronic collision.}
\label{fig:diagrams}
\end{figure}

\section{Formalism}
\label{form}

The typical diagrams for the heavy quark production in ultraperipheral collisions (UPC), where the photon-induced interactions are dominant, are represented in Fig. \ref{fig:diagrams}. The hadrons act as a source of 
almost real photons and the hadron-hadron cross section for the exclusive heavy quark photoproduction can be written in a factorized form, described using the equivalent photon approximation \cite{upc}. In the color dipole formalism \cite{nik},  the $\gamma h$ interaction can be expressed in terms of a  (color) dipole-hadron interaction and  the nonlinear effects in the QCD dynamics \cite{hdqcd} can be taken into account. In this formalism, the photon-hadron cross section for the exclusive heavy quark production is given by
\begin{eqnarray}
\sigma_{\gamma h \rightarrow  Q\bar{Q} h}(W_{\gamma h}) = \frac{1}{4} \int dz d^2\mathbf{r} \, |\Psi^{T}(z,\mathbf{r})|^2 \int d^2\mathbf{b}_{h} \left( \frac{d\sigma}{d^2\mathbf{b}_{h}} \right)^2\,,
\end{eqnarray}
where  $W_{\gamma h}$ is the photon -- hadron center -- of -- mass energy, $z$ is the photon momentum fraction carried by the quark, $\mathbf{r}$ is the transverse dipole separation and $\textbf{\textit{b}}_{h}$ is the impact parameter, given by the transverse distance  between the centers of the dipole  and the target. Moreover,  for a transversely polarized photon with $Q^2 = 0$ one has that the squared wave function $|\Psi^{T}(z,\mathbf{r})|^2$ is  given by \cite{nik}
\begin{eqnarray}   
|\Psi^{T}(z,\mathbf{r})|^2 &=& \frac{6\,\alpha_{em}\,e_{Q}^{2}}{(2\,\pi)^{2}}\left\{m_{Q}^{2}K_{0}^2  
(m_Q\, r)  
+ m_Q^{2}[z^{2}+(1-z)^{2}]  
K_{1}^2(m_Q\, r)\right\} \,,
\label{eq9}
\end{eqnarray}  
where $\alpha_{em}$ is the electromagnetic coupling constant, $e_Q$ is the fractional quark charge and $m_Q$ the mass of the heavy quark.
Furthermore, $x = 4m_Q^2 / W_{\gamma h}^2$ and the differential dipole-hadron cross section can be expressed by
\begin{eqnarray}
\frac{d\sigma}{d^2\mathbf{b}_{h}}= \,\, 2 \, {\cal N}_{h} 
(x,\textbf{\textit{r}},\textbf{\textit{b}}_{h}) ,
\end{eqnarray}
where  ${\cal N}_{h} (x,\textbf{\textit{r}},\textbf{\textit{b}}_{h})$ is the forward dipole-hadron scattering amplitude, which is dependent  on the modelling of the QCD dynamics at high energies.
As in our previous study \cite{nos_hq}, we will consider the  bCGC \cite{KMW} and IP-SAT \cite{ipsat2} models for the description of the dipole-proton scattering. Although these models differ in the treatment of the impact parameter dependence and/or of the linear and 
nonlinear regimes, 
both describe  quite well the high precision HERA data.  One has that in the  bCGC model,  the linear regime of the dipole-proton scattering amplitude is described by the solution of the BFKL dynamics near of the saturation line, which implies that 
${\cal N}_p \propto \rr^{2 \gamma_{eff}}$ with $\gamma_{eff} \le 1$. In contrast, the IP-SAT model predicts ${\cal N}_p \propto \rr^{2} \, xg(x,4/r^2)$ in the linear regime, where $xg$ is the gluon distribution of the target. On the other hand,  the saturation regime is described in the bCGC model by the  Levin-Tuchin law \cite{levin_tuchin}, while the IP-SAT predicts the saturation of $ {\cal N}_p$ at high energies and/or large dipoles, but the approach to this regime is not described by the Levin-Tuchin law.  In our analysis  we will assume  
for the bCGC model the set of parameters obtained in 
Ref. \cite{amir} by fitting the  HERA data on the reduced $ep$ cross sections. For the IP-SAT, we will assume the parameters obtained in Ref. \cite{ipsat_heikke}. In addition, we  will also present the predictions derived using the IPnonSAT model proposed in Ref. \cite{ipsat_heikke}, which is obtained 
disregarding the nonlinear corrections in the IP-SAT model. The comparison between the IP-SAT and IPnonSAT predictions will allow us to estimate the impact of the nonlinear corrections in the exclusive heavy quark photoproduction in hadronic collisions. 
For a nuclear target, we will assume that
the  dipole-nucleus scattering amplitude is given by the model 
proposed in Ref. \cite{armesto}, which is based on the Glauber-Gribov approach \cite{glauber,gribov,mueller}, being expressed by 
\begin{eqnarray}
 {\cal N}_{A}(x,\rr,\rb_A) = 1 - \exp \left[
-\frac{1}{2} \sigma_{dp}(x,r^{2}) \, T_{A}(\textbf{\textit{b}}_{A})
 \right] ,
 \label{Na_Glauber}
\end{eqnarray}
where
\begin{eqnarray}
 \sigma_{dp}(x,r^{2}) = 2 \int d^{2} \textbf{\textit{b}}_{p} \,\, {\cal N}_{p} 
(x,\textbf{\textit{r}},\textbf{\textit{b}}_{p}) ,
\end{eqnarray}
and  $T_{A}(b_A)$ is the nuclear thickness function 
which is typically obtained from the Woods-Saxon distribution 
for the nuclear density normalized to the atomic mass $A$, and $b_A$ is
the impact parameter of the dipole with respect to the nucleus center.
Although  such model  describes  the scarce existing  experimental data on the nuclear structure function \cite{erike}, future data from the Electron -- Ion Collider will be useful to constrain the description of the dipole -- nucleus scattering amplitude \cite{Boer:2011fh}. 
We will compute  $\mathcal{N}_A$ considering the bCGC, IP-SAT and IPnonSAT models for the dipole-proton scattering amplitude discussed before (For more details see e.g. Ref. \cite{nos_hq}).

\section{Results}
\label{res}

In what follows we will present our predictions for the rapidity distribution and cross sections for the exclusive charm and bottom photoproduction in $pp/pPb/PbPb$ collisions at the LHC and FCC energies. One has that
the differential cross section for the exclusive production of a heavy quark $Q\bar{Q}$ at rapidity $Y$ is given by
\begin{eqnarray}
\frac{d\sigma \,\left[h_1 + h_2 \rightarrow   h_1 + Q\bar{Q} + h_2\right]}{dY} = \left[n_{h_{1}} (\omega) \,\sigma_{\gamma h_2 \rightarrow Q\bar{Q} h_2}\left(W_{\gamma h_2}^2 \right)\right]_{\omega_L} + \left[n_{h_{2}} (\omega)\,\sigma_{\gamma h_1 \rightarrow  Q\bar{Q} h_1}\left(W_{\gamma h_1}^2 \right)\right]_{\omega_R}\,\,,
\label{dsigdy}
\end{eqnarray}
where $\omega_L \, (\propto e^{+Y})$ and $\omega_R \, (\propto e^{-Y})$ denote photons from the $h_1$ and $h_2$ hadrons, respectively. The center-of-mass energy for the photon-hadron interactions is given by  $W_{\gamma h} = \sqrt{4 \omega E}$, where $E = \sqrt{s}/2$ and $\sqrt{s}$ is the hadron-hadron center -- of -- mass energy. Moreover, $n(\omega)$ is the equivalent photon spectrum generated by  the 
hadronic source, which we will  assume to be described by the Drees-Zeppenfeld \cite{Dress} and the relativistic point-like charge \cite{upc} models for the case of a proton and a nucleus, respectively. The maximum photon energy can  be derived considering that the maximum possible momentum in the longitudinal direction is modified by the Lorentz factor, $\gamma_L$, due to the Lorentz contraction of the hadrons in that direction \cite{upc}. It implies $\omega_{\mbox{max}} \approx \gamma_L/R_h$ and, consequently, $W_{\gamma h}^{\mbox{max}} = \sqrt{2\,\omega_{\mbox{max}}\, \sqrt{s}}$. For the LHC, the maximum photon -- nucleon center -- of -- mass energy, $W_{\gamma h}^{\mbox{max}}$,  reached in $pp/pPb/PbPb$ collisions at $\sqrt{s} = 14 / 8.1 /5.5$ TeV is $8.4 /1.4/ 0.95$ TeV \cite{upc}. On the other hand, for the FCC we will reach  $W_{\gamma h}^{\mbox{max}} \approx 55/ 8.7/ 6.8$ TeV for 
$pp/pPb/PbPb$ collisions at $\sqrt{s} = 100 / 63 /  39$ TeV. Therefore,  LHC and FCC  probe a range of  photon -- hadron center -- of -- mass energies unexplored by HERA. 
In our calculations, we will assume $m_c = 1.27$ GeV and $m_b = 4.5$ GeV and  the phenomenological dipole-proton models discussed above, which describe the $ep$ HERA data, will be considered. The comparison between the IP-SAT, IPnonSAT and bCGC predictions will allow us  to  estimate the impact of the nonlinear effects, as well of the different descriptions of the transition between the linear and nonlinear regimes. In our analysis, we will assume that the events  can be  separated with a small experimental uncertainty, associated to e.g. the efficiency of charm and bottom tagging. Surely such aspect deserves a more detailed study in the future.

\begin{figure}[t]
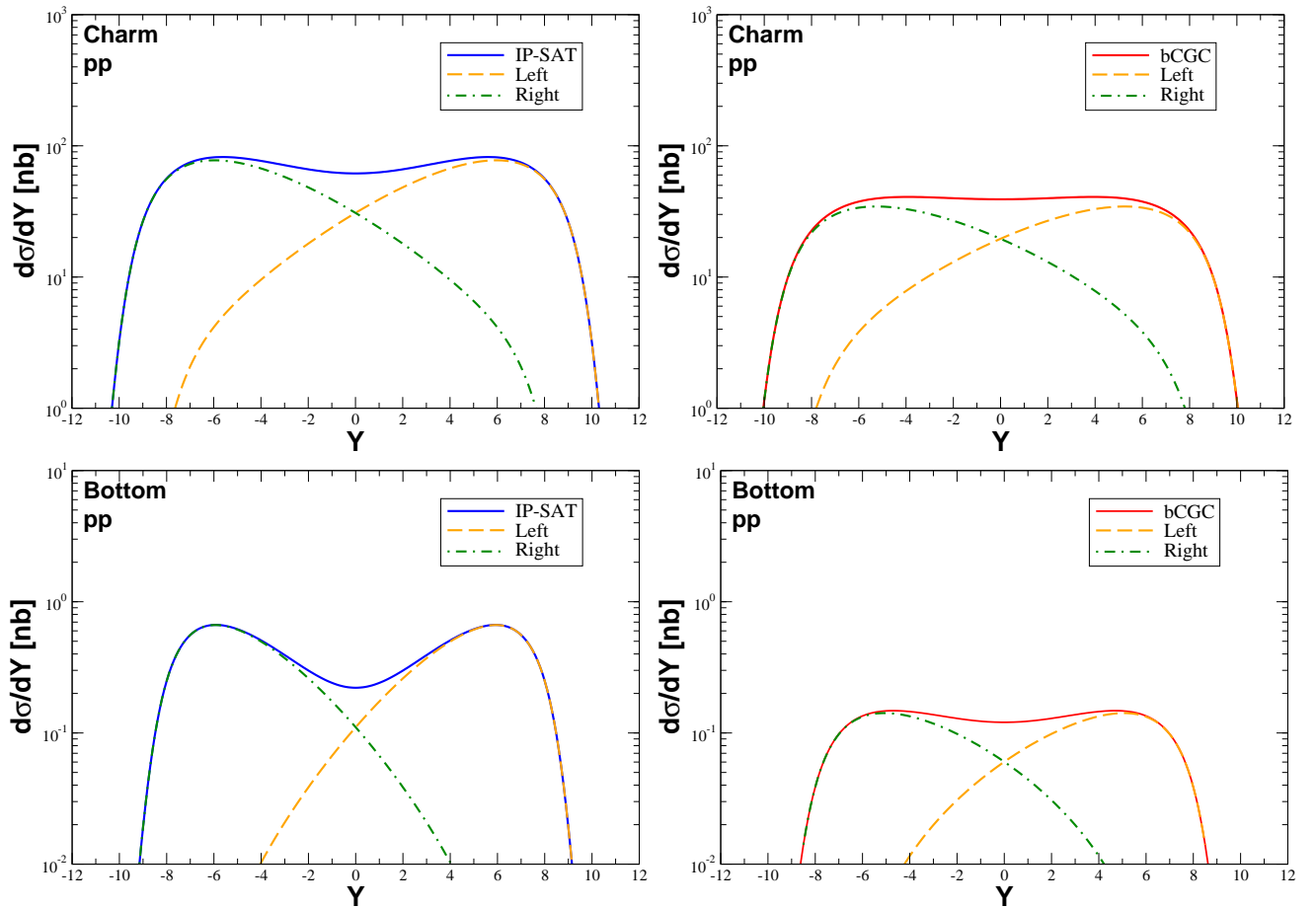

\begin{tabular}{cc}  
\includegraphics[scale=0.35]{dsigdydif_charm_pp100_comp.eps}
\includegraphics[scale=0.35]{dsigdydif_charm_pp100_comp_bcgc.eps} \\
\includegraphics[scale=0.35]{dsigdydif_bottom_pp100_comp.eps}
\includegraphics[scale=0.35]{dsigdydif_bottom_pp100_comp_bcgc.eps}
\end{tabular}
\caption{Rapidity distributions for the exclusive charm (upper panels) and bottom (lower panels) photoproduction in $pp$  collisions at the FCC ($\sqrt{s} = 100$ TeV).}
\label{fig:components}
\end{figure}

In Fig. \ref{fig:components} we present the IP-SAT and bCGC predictions for the rapidity distribution considering the exclusive  charm and bottom photoproduction in $pp$ collisions at the FCC ($\sqrt{s} = 100$ TeV). The contribution of both terms of Eq. (\ref{dsigdy}) are presented, as well as the sum of them is represented by the solid lines. The  first term in Eq. (\ref{dsigdy}), denoted ``Left" in the figures, is determined by the photon flux for a photon with a  energy $\omega \propto e^{Y}$ and the exclusive  heavy quark photoproduction cross section for a given photon -- proton center -- of -- mass energy $W_{\gamma p}$. While $\sigma_{\gamma p \rightarrow Q\bar{Q} p}$ increases with $W_{\gamma p}$, the  photon flux strongly decreases when the photon energy is of the order of $\omega_{\mbox{max}} \approx \gamma_L/R_p$, becoming almost zero for larger photon energies. As a consequence,  this contribution  increases with the rapidity up to a maximum and  becomes zero at very large $Y$. On the other hand, the second term in Eq. (\ref{dsigdy}), denoted ``Right" in the figures, increases for negative values of rapidity, since in this case $\omega \propto e^{-Y}$.  
For $pp$ collisions, the contributions of both terms are identical and symmetric in rapidity. Such behaviours are verified in Fig. \ref{fig:components}.
The increasing with rapidity is determined by the energy dependence of the exclusive heavy quark photoproduction cross section, being dependent on the dipole model considered. For the bottom production,  the IP-SAT model predicts a faster increasing with the energy than the bCGC one, which implies  that the sum of the ``Left" and ``Right" contributions have a smaller value  for central rapidity than for forward rapidities. Such behaviour is not present  in the bCGC result due to the milder increasing with rapidity predicted by this model. For the charm production, the IP-SAT and bCGC predictions for the energy dependence of $\sigma_{\gamma p \rightarrow c\bar{c} p}$   are similar, implying that both models predict a plateau for central rapidities.

\begin{figure}[t]
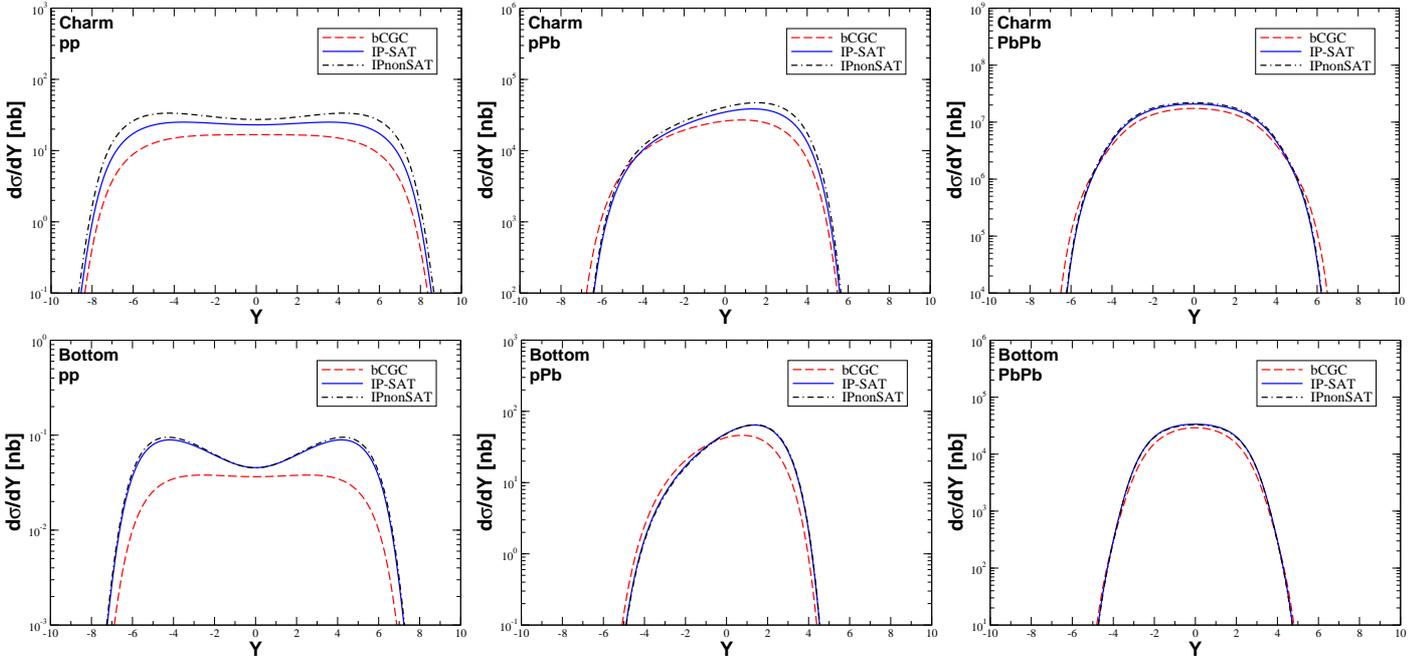

\begin{tabular}{ccc}  
\includegraphics[scale=0.25]{dsigdydif_charm_pp13.eps}
\includegraphics[scale=0.25]{dsigdydif_charm_pPb8100.eps}
\includegraphics[scale=0.25]{dsigdydif_charm_PbPb5020.eps} \\
\includegraphics[scale=0.25]{dsigdydif_bottom_pp13.eps}
\includegraphics[scale=0.25]{dsigdydif_bottom_pPb8100.eps}
\includegraphics[scale=0.25]{dsigdydif_bottom_PbPb5020.eps}
\end{tabular}
\caption{Rapidity distributions for the exclusive charm (upper panels) and bottom (lower panels) photoproduction in $pp$ ($\sqrt{s} = 13$ TeV), $pPb$ ($\sqrt{s} = 8.1$ TeV) and $PbPb$ ($\sqrt{s} = 5.02$ TeV) collisions at the LHC.}
\label{fig:comp_models_lhc}
\end{figure}

\begin{figure}[t]
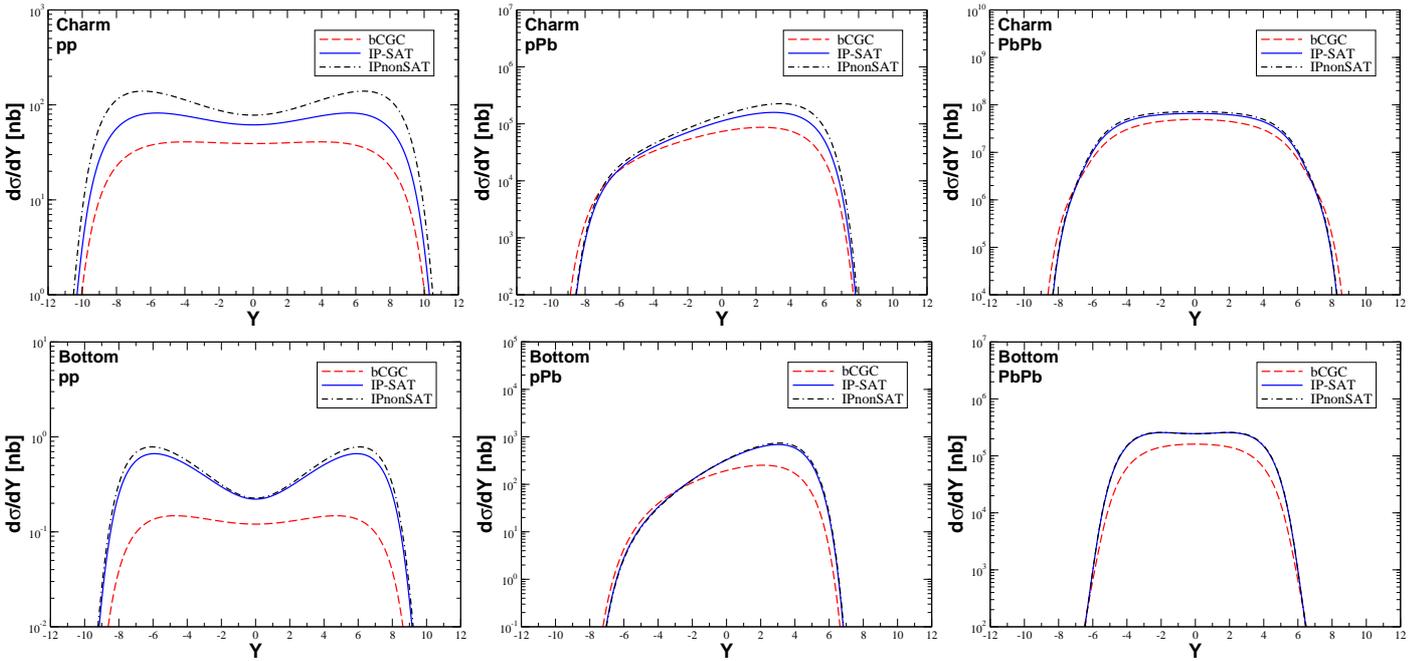

\begin{tabular}{ccc}  
\includegraphics[scale=0.25]{dsigdydif_charm_pp100.eps}
\includegraphics[scale=0.25]{dsigdydif_charm_pPb63.eps}
\includegraphics[scale=0.25]{dsigdydif_charm_PbPb39.eps} \\
\includegraphics[scale=0.25]{dsigdydif_bottom_pp100.eps}
\includegraphics[scale=0.25]{dsigdydif_bottom_pPb63.eps}
\includegraphics[scale=0.25]{dsigdydif_bottom_PbPb39.eps}
\end{tabular}
\caption{Rapidity distributions for the exclusive charm (upper panels) and bottom (lower panels) photoproduction in $pp$ ($\sqrt{s} = 100$ TeV), $pPb$ ($\sqrt{s} = 63$ TeV) and $PbPb$ ($\sqrt{s} = 39$ TeV) collisions at the FCC.}
\label{fig:comp_models_fcc}
\end{figure}

In  Figs. \ref{fig:comp_models_lhc} and \ref{fig:comp_models_fcc} we present a more detailed comparison between the IP-SAT, IPnonSAT and bCGC predictions for the charm (upper panels) and bottom (lower panels) photoproduction in $pp/pPb/PbPb$ collisions at the LHC and FCC energies, respectively. Some comments are in order. 
For $pPb$ and $PbPb$ collisions, the $Z^2$ factor, present in the nuclear photon flux, implies that the distributions   are larger.  For  $pPb$ collisions, we have that the distribution receives contributions of photon -- proton and photon -- nucleus interactions, with the photon -- proton contribution being larger. It implies that the rapidity distribution is asymmetric. Moreover, in this case, the behaviour of the distribution is determined by $\gamma p$ interactions and  the  rapidity directly determines the value of $x$ that is being probed: $x = 2m_Q e^{-Y}/\sqrt{s}$. For heavy quark production, the cross sections are dominated by the interaction of dipoles of size $r\approx 1/m_Q$ (See e.g. Ref. \cite{simone}), i.e.  the charm production  is dominated by  larger dipole sizes than for the bottom case.  As the impact of the nonlinear effects increases with the size of the dipole, we expect a larger contribution of the nonlinear effects in the case of charm production. Moreover, for the larger photon-hadron center-of-mass energies achieved at the FCC in comparison to the LHC, we also expect a larger contribution of the nonlinear effects since the impact of these effects increases at smaller values of $x$. Such expectations are confirmed in the results presented in Figs. \ref{fig:comp_models_lhc} and \ref{fig:comp_models_fcc}. We have that the difference between the IP-SAT and IPnonSAT predictions is negligible for bottom production, large for charm production and increases with the energy.
One can also observe a large difference between the IP-SAT and bCGC predictions, which is directly related to distinct descriptions of the linear and nonlinear regimes, as well as for the transition between these regimes. In particular, the large difference observed in the predictions for the bottom production in $pp$ collisions is explained by the distinct treatments of the linear regime, which is dominated by the interaction of very small dipoles ($r\approx 1/m_b$) with the proton. Another way to observe these distinct treatments of the linear and nonlinear regimes present in the phenomenological models is the  analysis of the results for the charm production in $pPb$ collisions. As pointed above, in this case the behaviour of the rapidity distribution is directly associated to  the value that is being probed in the scattering amplitude. Therefore, for negative (positive) values of rapidity we are probing  ${\cal{N}}_p$ at large (small)-$x$. One has that bCGC prediction is larger than the IP-SAT one in the linear regime and smaller in the saturation regime.    
For $PbPb$ collisions, this difference is smaller, which is directly associated to the fact that we are using the Glauber -- Gribov model \cite{armesto} to describe the dipole-nucleus scattering, with the bCGC and IP-SAT only affecting the argument of the exponential. In addition, the rapidity distributions are narrower in comparison to those for $pp$ collisions.
This behaviour is associated to the fact that the Lorentz factor $\gamma_L$ is smaller for a $Pb$ beam and $R_{Pb} > R_p$. Consequently, the value of  $\omega_{\mbox{max}} \approx \gamma_L/R_h$, where the rapidity distribution strongly decreases,  is  smaller  for  $PbPb$ collisions in comparison to $pp$ one.

In Tables \ref{tab:charm} and \ref{tab:bottom} we present the corresponding predictions for the charm and bottom cross sections, respectively, considering the rapidity ranges probed  by the CMS ($-2.5 \le Y \le +2.5$) and LHCb ($2.0 \le Y \le 4.5$) detectors.  
One has that the predictions for the LHCb range are approximately a factor $\ge 2$ smaller than those for the  CMS kinematical range. 
We predict large values for the cross sections, in particular for the charm production in $PbPb$ collisions and FCC energies. 
One important question is the number of events associated to these cross sections. For the typical $pp$ collisions at the LHC, the integrated luminosity per year is expected to be $\approx 1$ fb$^{-1}$, which implies that the number of events per year will be larger than $39 \, (0.1) \times 10^6$ for charm (bottom) production.  For the high -- luminosity LHC \cite{fcc}, these numbers will be enhanced by a factor 350. 
Finally, for the FCC, where the integrated luminosity per year is expected to be $\ge 1000$ fb$^{-1}$,  the associated number of events will be $\ge 10^9$.
For $PbPb$ collisions, the expected integrated luminosities per year for the next run of the LHC and for the FCC are 3 nb$^{-1}$ and 110 nb$^{-1}$, respectively. Consequently, we predict that the number of events per year associated to charm (bottom) production in these collisions will be larger than $10^7 \, (10^4)$.
These large numbers for the event rates indicate that  a future measurement of the exclusive charm and bottom photoproduction in hadronic collisions is, in principle, feasible and that the analysis of this observable can be useful to constrain the description of the QCD dynamics at high energies.
 Based on these large rates, we believe that the forthcoming analysis at the LHC and FCC will also allow to perform the analysis of the differential distributions needed to constrain the elliptic component of the gluon Wigner distribution.

\begin{table}[t]
\centering
\begin{tabular}{|c|c|c|c|c|}\hline \hline
      &   {\bf Rapidity range} &                       {\bf bCGC}               &     {\bf IP-SAT}        & {\bf IPnonSAT}                                \\ \hline \hline        
{\bf pp ($\sqrt{s} = 13$ TeV)} & $-2.5 < Y < 2.5$   & 83.2 nb                  & 117.9 nb               & 142.7 nb \\                       
  & $2 < Y < 4.5$                                   & 39.3 nb                  & 61.9 nb                & 80.2 nb \\
   \hline        
{\bf pp ($\sqrt{s} = 100$ TeV)} & $-2.5 < Y < 2.5$   & 197.6 nb                 & 320.0 nb               & 415.0 nb \\                       
  & $2 < Y < 4.5$                                    & 101.2 nb                 & 181.6 nb               & 257.1 nb \\
\hline
\hline
{\bf pPb ($\sqrt{s} = 8.1$ TeV)} & $-2.5 < Y < 2.5$ & 12.0$\times10^{4}$ nb     & 16.0$\times10^{4}$ nb   & 19.0$\times10^{5}$ nb \\
  &  $2 < Y < 4.5$                                  & 3.7$\times 10^{4}$ nb    & 6.0$\times 10^{4}$ nb  & 7.7$\times 10^{4}$ nb \\
\hline
{\bf pPb ($\sqrt{s} = 63$ TeV)} & $-2.5 < Y < 2.5$   & 3.6$\times10^{5}$ nb     & 5.6$\times10^{5}$ nb   & 7.1$\times10^{5}$ nb \\
  &  $2 < Y < 4.5$                                   & 2.0$\times 10^{5}$ nb    & 3.8$\times 10^{5}$ nb  & 5.5$\times 10^{5}$ nb \\
\hline
\hline
{\bf PbPb ($\sqrt{s} = 5.02$ TeV)} &$-2.5 < Y < 2.5$& 7.7$\times10^{7}$ nb     & 9.3$\times10^{7}$ nb   & 9.9$\times 10^{7}$ nb \\
  &  $2 < Y < 4.5$                                  & 1.9$\times 10^{7}$ nb    & 2.4$\times 10^{7}$ nb  & 2.5$\times 10^{7}$ nb \\
  \hline
{\bf PbPb ($\sqrt{s} = 39$ TeV)} &$-2.5 < Y < 2.5$   & 2.3$\times10^{8}$ nb     & 3.2$\times10^{8}$ nb   & 3.5$\times 10^{8}$ nb \\
  &  $2 < Y < 4.5$                                   & 0.9$\times 10^{8}$ nb    & 1.3$\times 10^{8}$ nb  & 1.4$\times 10^{8}$ nb \\
\hline  
\end{tabular} 
\caption{Cross sections for the exclusive charm photoproduction in $pp/pPb/PbPb$ collisions at the LHC and FCC energies considering two rapidity ranges.}
\label{tab:charm}
\end{table}

\begin{table}[t]
\centering
\begin{tabular}{|c|c|c|c|c|}\hline \hline
      &   {\bf Rapidity range} &                       {\bf bCGC}               &     {\bf IP-SAT}        & {\bf IPnonSAT}                                \\ \hline \hline        
{\bf pp ($\sqrt{s} = 13$ TeV)} & $-2.5 < Y < 2.5$   & 0.2 nb                   & 0.3 nb                 & 0.3 nb \\                       
  & $2 < Y < 4.5$                                   & 0.1 nb                   & 0.2 nb                 & 0.2 nb \\
  \hline        
{\bf pp ($\sqrt{s} = 100$ TeV)} & $-2.5 < Y < 2.5$   & 0.6 nb                  & 1.3 nb                 & 1.4 nb \\                       
  & $2 < Y < 4.5$                                    & 0.4 nb                  & 1.1 nb                 & 1.2 nb \\
\hline
\hline
{\bf pPb ($\sqrt{s} = 8.1$ TeV)} & $-2.5 < Y < 2.5$ & 176.1 nb                 & 217.5 nb               & 216.5 nb \\
  &  $2 < Y < 4.5$                                  & 30.6 nb                  & 57.2 nb                & 58.7 nb \\
\hline
{\bf pPb ($\sqrt{s} = 63$ TeV)} & $-2.5 < Y < 2.5$   & 937.5 nb                & 1750.2 nb              & 1804.2 nb \\
  &  $2 < Y < 4.5$                                   & 527.1 nb                & 1572.1 nb              & 1696.1 nb \\
\hline
\hline               
{\bf PbPb ($\sqrt{s} = 5.02$ TeV)} &$-2.5 < Y < 2.5$& 10.1$\times10^{4}$ nb     & 13.4$\times10^{4}$ nb   & 13.3$\times 10^{4}$ nb \\
  &  $2 < Y < 4.5$                                  & 1.1$\times 10^{4}$ nb    & 1.5$\times 10^{4}$ nb  & 1.4$\times 10^{4}$ nb \\
\hline
{\bf PbPb ($\sqrt{s} = 39$ TeV)} &$-2.5 < Y < 2.5$   & 7.6$\times10^{5}$ nb     & 12.6$\times10^{5}$ nb   & 12.7$\times 10^{5}$ nb \\
  &  $2 < Y < 4.5$                                   & 2.4$\times 10^{5}$ nb    & 5.0$\times 10^{5}$ nb  & 5.1$\times 10^{5}$ nb \\
\hline  
\end{tabular} 
\caption{Cross sections for the exclusive bottom photoproduction in $pp/pPb/PbPb$ collisions at the LHC and FCC energies considering two rapidity ranges.}
\label{tab:bottom}
\end{table}

\begin{figure}[t]
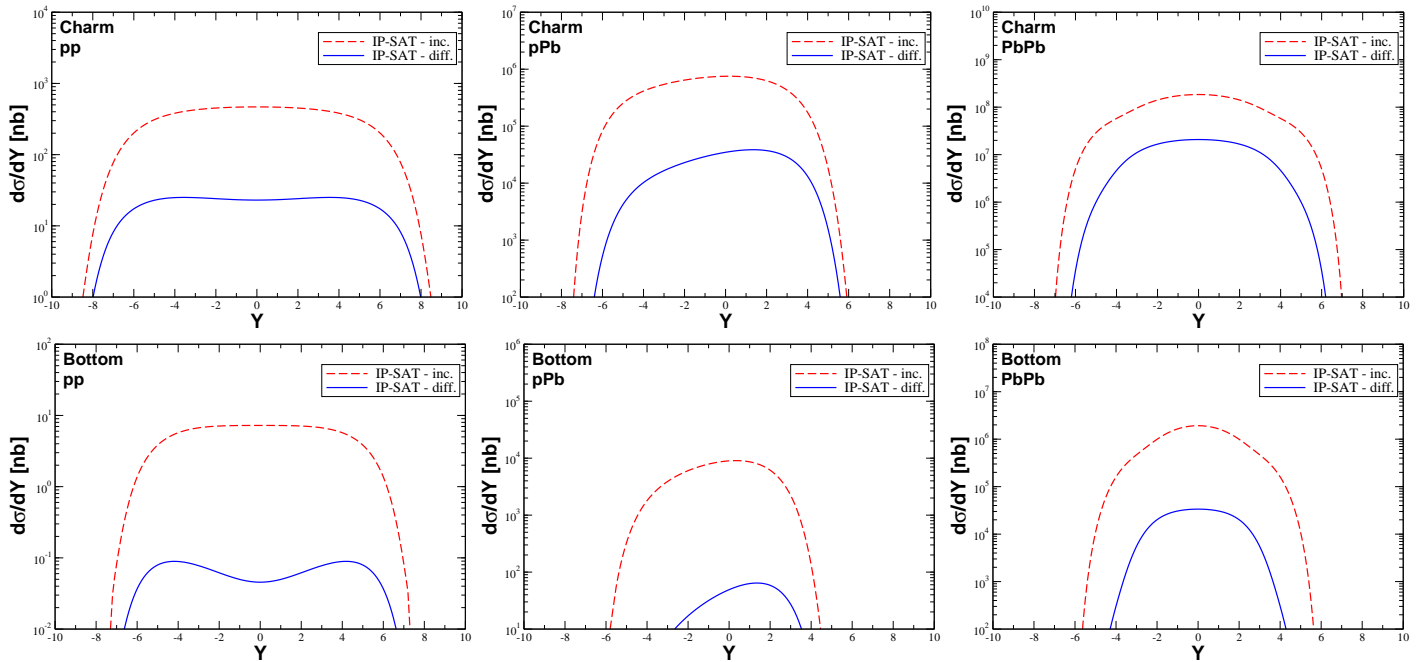

\begin{tabular}{ccc}  
\includegraphics[scale=0.25]{dsigdy_charm_inc_diff_pp13.eps}
\includegraphics[scale=0.25]{dsigdy_charm_inc_diff_pPb8100.eps}
\includegraphics[scale=0.25]{dsigdy_charm_inc_diff_PbPb5020.eps} \\
\includegraphics[scale=0.25]{dsigdy_bottom_inc_diff_pp13.eps}
\includegraphics[scale=0.25]{dsigdy_bottom_inc_diff_pPb8100.eps}
\includegraphics[scale=0.25]{dsigdy_bottom_inc_diff_PbPb5020.eps}
\end{tabular}
\caption{Comparison between the rapidity distributions for the inclusive and exclusive charm (upper panels) and bottom (lower panels) photoproduction in $pp$ ($\sqrt{s} = 13$ TeV), $pPb$ ($\sqrt{s} = 8.1$ TeV) and $PbPb$ ($\sqrt{s} = 5.02$ TeV) collisions at the LHC.}
\label{fig:comp_lhc}
\end{figure}

\begin{figure}[t]
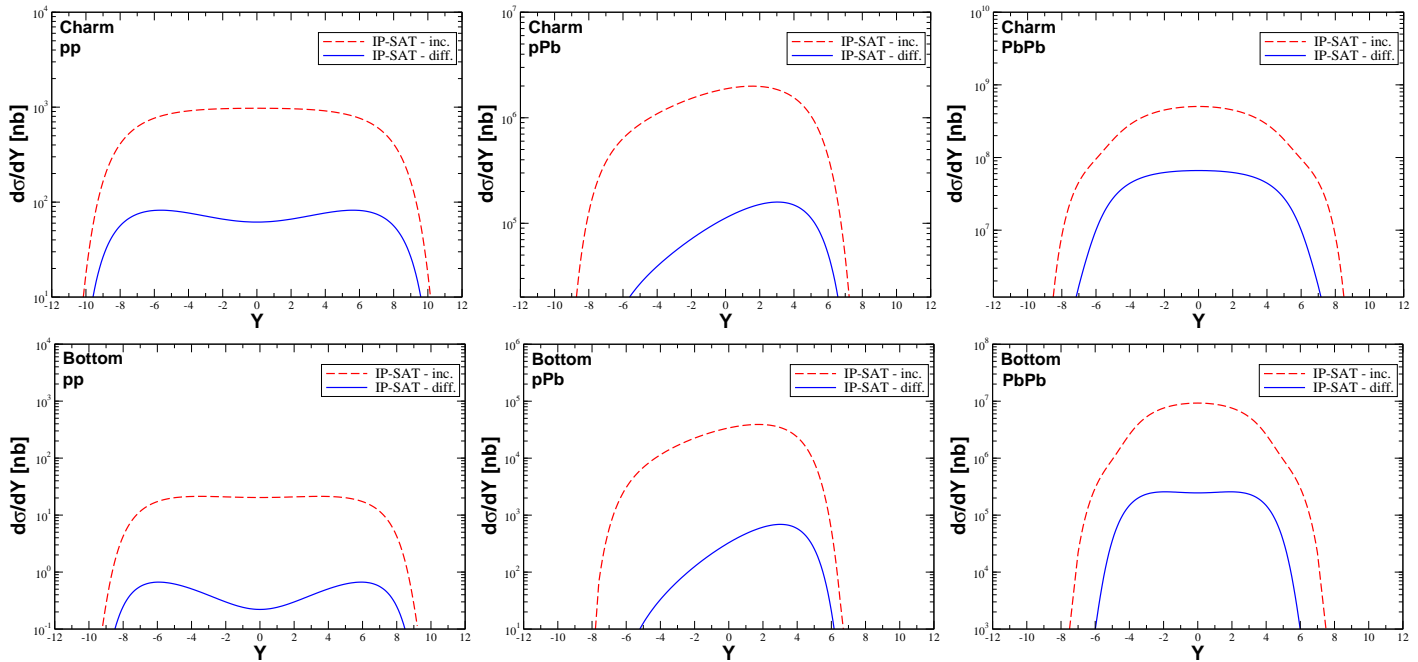

\begin{tabular}{ccc}  
\includegraphics[scale=0.25]{dsigdy_charm_inc_diff_pp100.eps}
\includegraphics[scale=0.25]{dsigdy_charm_inc_diff_pPb63.eps}
\includegraphics[scale=0.25]{dsigdy_charm_inc_diff_PbPb39.eps} \\
\includegraphics[scale=0.25]{dsigdy_bottom_inc_diff_pp100.eps}
\includegraphics[scale=0.25]{dsigdy_bottom_inc_diff_pPb63.eps}
\includegraphics[scale=0.25]{dsigdy_bottom_inc_diff_PbPb39.eps}
\end{tabular}
\caption{Comparison between the rapidity distributions for the inclusive and exclusive charm (upper panels) and bottom (lower panels) photoproduction in $pp$ ($\sqrt{s} = 100$ TeV), $pPb$ ($\sqrt{s} = 63$ TeV) and $PbPb$ ($\sqrt{s} = 39$ TeV) collisions at the FCC.}
\label{fig:comp_fcc}
\end{figure}

Finally, let's compare the predictions for the charm and bottom photoproduction in inclusive and exclusive processes, derived using the IP-SAT model. In contrast to the exclusive case, in inclusive interactions one of the incident hadrons fragments and the photon-hadron cross section is linearly proportional to ${d\sigma}/{d^2\mathbf{b}_{h}}$  (For details see, e.g. \cite{nos_hq}).  In Figs. \ref{fig:comp_lhc} and \ref{fig:comp_fcc} we present our results for $pp/pPb/PbPb$ collisions at the LHC and FCC energies, respectively. The results for charm (bottom) production are presented in the upper (lower) panels.  In the case of charm production  in $pp$ and $pPb$ collisions at the LHC, we have that the exclusive prediction is a factor ${\cal{O}}(20)$ smaller than the inclusive one for midrapidities. On the other hand, for $PbPb$ collisions, this factor is ${\cal{O}}(10)$. For the FCC we have that the corresponding factors are of order of 15/18/8 for $pp/pPb/PbPb$ collisions, respectively. 
Such results indicate that the exclusive charm production is not strongly suppressed in comparison to the inclusive case. 
In contrast, we have that exclusive bottom photoproduction is a factor ${\cal{O}}(100)$ [${\cal{O}}(80)$] smaller than the inclusive one for the LHC [FCC] energies, which implies that the experimental separation of these events in the future experimental analysis will be a more difficult task. It is important to emphasize that 
the free parameters present in the color dipole formalism have been    
constrained by the HERA data, which implies that our predictions  are parameter free. Moreover, 
as  the inclusive and exclusive photoproduction in hadronic collisions are determined by the same quantities, i.e. the photon wave function and the color-dipole amplitude,  a future measurement of both processes will be an important test of the universality of the color dipole formalism as well of the treatment of the nonlinear corrections to the QCD dynamics. In addition, as the Bjorken -- $x$ range probed at the FCC is beyond that probed at HERA, our predictions are based on the extrapolation of these models for a new kinematical range, where higher  -- order corrections can become important \cite{lappi}. In principle, the magnitude of such corrections can also be constrained by the future FCC data. 

\section{Summary}
\label{conc}
In this paper we have computed the rapidity distributions and cross sections for the exclusive charm and bottom photoproduction in $pp/pPb/PbPb$ collisions at the LHC and FCC energies. Our study was motivated by the possibility of use this process to constrain the elliptic component of the gluon Wigner distribution by the analysis of the 
differential distribution in the relative quark-antiquark momentum for distinct values  of the momentum transfer. Our predictions  indicate that the expected number of events is very large, in particular for charm production  at the FCC. Therefore, we strongly recommend a future experimental analysis of this process in order to improve our understanding of the QCD dynamics and to access the gluon Wigner distribution.

%(see \cite{contreras} for a recent review)

\section*{Acknowledgements}
VPG  would like to express a special thanks to the Mainz Institute for Theoretical Physics (MITP) of the Cluster of Excellence PRISMA+ (Project ID 39083149) for its hospitality and support.   
This work was partially financed by the Brazilian funding agencies CAPES, CNPq,  FAPERGS and INCT-FNA (process number 464898/2014-5).

%%%%%%%%%%%%%%%%%%%%%%%%%%%%%%%%%%%%%%%%%%%%%%%%%%%%%%%%%%%%%%%%%%%%%%%%%%%%%%%%

%%%%%%%%%%%%%%%%%%%%%%%%%%%%%%%%%%%%%%%%%%%%%%%%%%%%%%%%%%%%%%%%%%%%%%%%%%%%%%%%

\end{document}